\documentclass{article} 
\usepackage{iclr2024_conference,times}

\usepackage{amsmath,amsfonts,bm}

\def\eqref#1{equation~\ref{#1}}

\def\1{\bm{1}}

\def\rx{{\textnormal{x}}}
\def\ry{{\textnormal{y}}}

\DeclareMathAlphabet{\mathsfit}{\encodingdefault}{\sfdefault}{m}{sl}
\SetMathAlphabet{\mathsfit}{bold}{\encodingdefault}{\sfdefault}{bx}{n}

\newcommand{\E}{\mathbb{E}}

\newcommand{\sigmoid}{\sigma}

\DeclareMathOperator*{\argmax}{arg\,max}

\usepackage{hyperref}
\usepackage{url}
\usepackage{graphicx}
\usepackage{subfigure}
\usepackage{wrapfig}
\title{Preference optimization of protein language models as a multi-objective binder design paradigm}

\author{Pouria Mistani \thanks{Corresponding author.}\\
Aikium Inc.\\
Berkeley, CA 94704, USA \\
\texttt{pouria@aikium.com} \\
\And
Venkatesh Mysore \\
Aikium Inc.\\
Berkeley, CA 94704, USA \\
\texttt{venkatesh@aikium.com} \\
}

\iclrfinalcopy 
\begin{document}

\maketitle

\begin{abstract}
We present a multi-objective binder design paradigm based on instruction fine-tuning and direct preference optimization (DPO) of autoregressive protein language models (pLMs). Multiple design objectives are encoded in the language model through direct optimization on expert curated preference sequence datasets comprising preferred and dispreferred distributions. We show the proposed alignment strategy enables ProtGPT2 to effectively design binders conditioned on specified receptors and a drug developability criterion. Generated binder samples demonstrate median isoelectric point (pI) improvements by $17\%-60\%$.

\end{abstract}

\section{Introduction}
\label{sec:Intro}
Peptides are an important class of biomolecules comprised of short strands of up to 50 amino acids. Designing peptide binders to specific protein targets with desirable therapeutic properties is a central problem in drug discovery \citep{FOSGERAU2015122}. 
Beyond optimizing for binding affinity, peptide drug development processes require satisfying numerous other constraints imposed by physicochemical properties, their formulation characteristics, and pharmacodynamic influences on human subjects, among others. Recent progress in generative artificial intelligence has inspired novel strategies for designing protein binders, rooted in either structure-based or sequence-based protein representations. Filtering for designs that satisfy other objectives is performed \textit{post facto}. A computational approach for generating peptides likely to satisfy multiple property constraints directly inferred from positive and negative examples is lacking; this work attempts to fill that lacuna.

Extending large language models (LLMs) for natural language processing (NLP) to biological sequences, protein language models (pLMs) are pre-trained on large scale evolutionary sequence data. Prominent foundation models include ESM2 \citep{lin2022language} that is a BERT-style encoder transformer model, ProtGPT2 \citep{ferruz2022protgpt2} and ProGen2 \citep{progen2} that are GPT-style decoder transformer models, and ProtT5 \citep{elnaggar2021prottrans} that is an encoder-decoder transformer model. These have been adapted for binder design models by fine-tuning these foundation models with different strategies. Examples of binder design models include
PepMLM \citep{chen2023pepmlm}, DiffPALM \citep{lupo2023pairing}, IgLM \citep{shuai2023iglm}, and pAbT5 \citep{chu2023generative}. These binder design models can be categorized by their formulation of the binder design problem as different NLP tasks. Three NLP tasks can be identified, (i) \textit{text infilling} with masked language modeling such as PepMLM that is an encoder transformer based on fine-tuning ESM2 to reconstruct the fully masked binder region in a protein-peptide conjugated chain, (ii) \textit{text generation} task with causal language models such as IgLM and ProtGPT2 that are decoder transformers based on GPT-2 \citep{radford2019language}, and (iii) \textit{machine translation} task using sequence-to-sequence language modeling such as pAbT5 that generates the heavy/light chain given their chain pairing partner in antibody sequences by fine-tuning the decoder stage in ProtT5.



On the other hand, structure-based models attempt to reason about binding and other properties by carefully constructing in three dimensions bound poses of the protein-peptide complex \citep{chang2024}. Despite being hindered by limited availability of solved protein-peptide complex structures and the computational cost of molecular dynamics simulations, good progress has been made in structure-informed peptide design\citep{kosugi2022,goudy2023,bryant2023}. More recently, diffusion models in the structure space \citep{anand2022protein,watson2023novo} as well as the sequence space \citep{alamdari2023protein,gruver2023protein}
are being adapted for peptide design \citep{xie2023,wang2024}.

Drug discovery and development is a multi-objective optimization process. Beyond binding affinity, numerous other factors need to be considered for therapeutic development such as expressibility, synthesizability, stability, immunogenicity, solubility and bioavailability. Our goal is to develop a framework for binder design beyond binding affinity, where downstream properties and experimental heuristics from human experts can be readily incorporated in a multi-objective optimization framework. Interestingly, techniques such as Reinforcement Learning from Human Feedback (RLHF) \citep{christiano2017deep,bai2022constitutional} and Direct Preference Optimization (DPO) \citep{rafailov2023direct} can instill desired behaviors in the responses generated by large language models; \textit{e.g.}, see \citep{park2023preference} for unconditional small molecule generation. 
In this study, we optimize autoregressive pLMs to capture diverse preferred and undesired protein sequence distributions, while conditioned on target receptor sequences. This approach enables development of computational frameworks for multi-objective drug design.
We use positive and negative data distributions for protein-peptide binding affinity as well as peptide isoelectric points (pI) to show that the model can learn to generate novel sequences that simultaneously respect these different objectives. 


\section{Methods}
We propose an alignment method to transform pre-trained unconditional protein sequence models ($\displaystyle p(s)$), that autoregressively sample sequences ($s$) from underlying data distribution ($\mathcal{D}$), to conditional probability models ($\displaystyle p(s|r;c)$) that given a target receptor ($r$) sample binders that satisfy constraints ($c$) encoded by preference datasets compiled from experiments and domain experts. 

\begin{figure}[h]
\begin{center}
\includegraphics[width=0.8\linewidth]{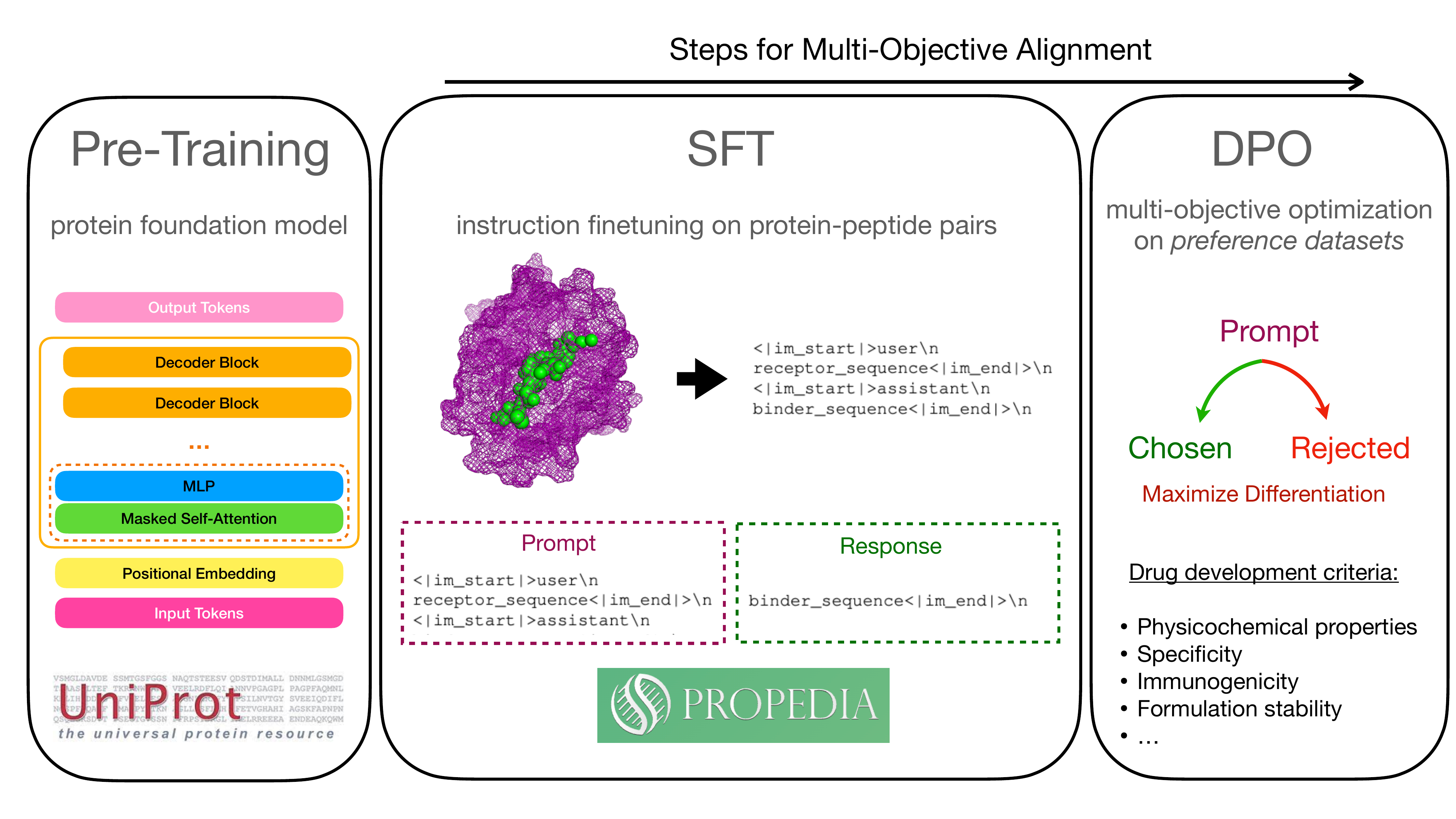}
\end{center}\vspace{-0.2in}
\caption{Alignment method for multi-objective optimization of favorable binders}
\label{fig:method}
\end{figure}
We perform a two step instruction fine-tuning, as in \citep{ouyang2022training}, see figure \ref{fig:method}: (i) we instill receptor-binder \textit{chat-templates} given in table \ref{tab:chat-template_chatml} through supervised fine-tuning (SFT), and (ii) we optimize the fine-tuned model to promote preferred binders over dispreferred ones. In this work, we curate preference datasets to induce high specificity binders with favorable isoelectric point values (\textit{i.e.}, pH at which the peptide is electrically neutral). Specifically, for ease of illustration, we demonstrate how to nudge the model to generate peptides with a higher pI. 


\subsection{Supervised fine-tuning (SFT) for protein-peptide binders}
We fine-tuned ProtGPT2 for instruction tasks following the OpenAI chatML template \citep{openai-chatml}, see table \ref{tab:chat-template_chatml}. We fine-tuned all linear layers (including embedding layers, attention blocks and MLPs) with QLoRA optimization \citep{dettmers2023qlora} for 24 epochs; see appendix \ref{app:trainparams} for more details. The model learns to generate binders when prompted by the generation template for a given receptor. The sequences follow FASTA convention of inserting next-line special character `\texttt{\textbackslash n}' after every $60$ residues. Note that ProtGPT2 uses the Byte Pair Encoding (BPE) tokenizer \citep{sennrich2015neural} with $50,257$ vocabulary size (recall we added two extra tokens) that contains higher order oligomers up to 9 residues long. This is in contrast to common pLMs such as ESM2 with only $33$ vocabulary size. Supporting more tokens primes the foundation model for more complex design tasks such as AI design agents from textual descriptions. 
\begin{table}[t]
\caption{OpenAI ChatML template for receptor-binder instruction fine-tuning}\vspace{-0.05in}
\label{tab:chat-template_chatml}
\begin{center}
\begin{tabular}{lr}
\multicolumn{1}{l}{\bf Message template} & \multicolumn{1}{r}{\bf Message segments}
\\ \hline \\
\begin{minipage}{2in}
\begin{tiny}
\begin{verbatim}
<|im_start|>user\n                                   
receptor_sequence<|im_end|>\n
<|im_start|>assistant\n                   
binder_sequence<|im_end|>\n        
\end{verbatim}
\end{tiny}
\end{minipage} & \begin{minipage}{2in}
\begin{tiny}
\begin{verbatim}
|                         |
| -> (generation prompt)  |
|                         | -> (SFT message)
  -> (completion)         |
\end{verbatim}
\end{tiny}
\end{minipage} 
\end{tabular}
\end{center}\vspace{-0.2in}
\end{table}


Fine-tuning is performed using a causal language modeling objective. Each receptor-binder sequence pair is represented in the format of table \ref{tab:chat-template_chatml}, then tokenized into a set of symbols $s=(t_1,\cdots, t_n)$. Assuming probability of next token depends on preceding tokens, the total probability of a sequence pair, $s^{(k)}$, is $\displaystyle p(s^{(k)})=\Pi_{i=1}^{n}p(t_i|t_1,\cdots,t_{i-1})$. Therefore, we estimate a pLM to predict conditional probabilities, $p\sim \pi_\theta(t_i|t_{<i})$, by minimizing the model negative log-likelihoods:
\begin{align*}
    \mathcal{L}_{SFT}(\pi_\theta)=\E_{(t_i\in s; s\sim\mathcal{D})}\big[-\log \pi_\theta(t_i|t_{<i})\big]
\end{align*}

\subsection{Direct preference optimization (DPO) for multi-objective design}
Let $\displaystyle \pi_\theta$ be the pre-trained protein language model. In the first phase, the model is subjected to supervised fine-tuning (SFT) for the instruction task using a high quality receptor-binder dataset $\displaystyle (x,y)$, to obtain a reference model $\displaystyle \pi_{\rm ref}=\pi^{\rm SFT}$. In the second phase, the reference model is prompted with target proteins, $x$, to sample binder pairs, $\displaystyle y_w, y_l\sim \pi_{ref}(\cdot | x)$, where $y_w$ is preferred and $y_l$ is dispreferred by some criteria. In practice, we use carefully curated offline preference datasets $\displaystyle \mathcal{D}=\{x^{(i)}, y^{(i)}_w, y^{(i)}_l\}_{i=1}^N$. The preferences are assumed to be sampled from the Bradley-Terry (BT) \citep{bradley1952rank} reward model $\displaystyle r^\ast(y,x)$:
\begin{align}
    p^\ast(y_1\succ y_2 | x) = \sigmoid\big(r^\ast(x,y_1) - r^\ast(x,y_2)\big) \label{eq:BT} 
\end{align}
In RLHF, a parameterized reward model $r_\phi(y,x)$ is used to provide feedback for optimizing the language model. The reward model is explicitly estimated using the negative log-likelihood loss $\displaystyle  \E_{(\rx, \ry_w, \ry_l)\sim \mathcal{D}} [- \log \sigmoid(r_\phi(x,y_w) - r_\phi(x, y_l))]$ on the preference data, which is then used to optimize the optimal policy $\displaystyle \pi^\ast=\argmax_{\pi_\theta}\E_{x\in\mathcal{D},y\in\pi_\theta}[r_\phi(x,y)]-\beta D_{KL}\big( \pi_\theta(y|x)||\pi_{ref}(y|x)\big)$. This is commonly achieved using the PPO algorithm \citep{schulman2017proximal}. In DPO, the reward function is explicitly expressed in terms of its optimal policy $\displaystyle r^\ast(x,y)=\beta\log\frac{\pi^\ast(y|x)}{\pi_{ref}(y|x)}+\beta\log Z(x)$. Thus, the optimal model can be expressed only in terms of the optimal policy and the reference policy, bypassing the need for estimating the reward model. Therefore, the DPO loss is directly minimizing the negative log-likelihood of the preference model in equation \ref{eq:BT}:
\begin{align}
    \displaystyle \mathcal{L}_{DPO}(\pi_\theta; \pi_{ref})=\E_{(\rx,\ry_w,\ry_l)\sim\mathcal{D}}\bigg[-\log \sigmoid \bigg( \beta \log \frac{\pi_\theta(y_w|x)}{\pi_{ref}(y_w|x)} - \beta\log \frac{\pi_\theta(y_l|x)}{\pi_{ref}(y_l|x)}\bigg) \bigg]
\end{align}
where $\displaystyle\beta$ determines information retention from the reference model.

\subsection{Preference datasets: specificity \& isoelectric points}
For each target protein we can provide several preference datasets. The preference datasets follow the format given in table \ref{tab:preference-data-format}.
\begin{table}[t]
\caption{Format of preference data for DPO}\vspace{-0.1in}
\label{tab:preference-data-format}
\begin{center}
\begin{tabular}{lccr}
\multicolumn{1}{l}{\bf Prompt}  &\multicolumn{1}{c}{\bf Chosen} &\multicolumn{1}{c}{\bf Rejected} & \multicolumn{1}{r}{\bf Criteria}
\\ \hline 
    \begin{minipage}{2in}
    \begin{tiny}
\begin{verbatim}

<|im_start|>user\n
LRGLSEDTLEQLYALGFNQ...<|im_end|>\n
<im_start>assistant\n

\end{verbatim}
    \end{tiny}
    \end{minipage}
& \begin{minipage}{1.1in}
    \begin{tiny}
    \begin{verbatim}
    
  TGVALTPPS<|im_end|>\n
  
    \end{verbatim}
    \end{tiny}
    \end{minipage} & \begin{minipage}{1.1in}
    \begin{tiny}
    \begin{verbatim}
    
    CRGCX<|im_end|>\n
    
    \end{verbatim}
    \end{tiny}
    \end{minipage} & \begin{small}affinity/specificity\end{small} \\
    \hline 
    \begin{minipage}{2in}
    \begin{tiny}
\begin{verbatim}

<|im_start|>user\n
LRGLSEDTLEQLYALGFNQ...<|im_end|>\n
<im_start>assistant\n

\end{verbatim}
    \end{tiny}
    \end{minipage}
& \begin{minipage}{1.1in}
    \begin{tiny}
    \begin{verbatim}
  TGVALTPPS<|im_end|>\n
    \end{verbatim}
    \end{tiny}
    \end{minipage} & \begin{minipage}{1.1in}
    \begin{tiny}
    \begin{verbatim}
  TGVDLTEPS<|im_end|>\n
    \end{verbatim}
    \end{tiny}
    \end{minipage} & \begin{small}isoelectric point\end{small} \\ \hline
\end{tabular}
\end{center}
\vspace{-0.2in}
\end{table}
Rejection criteria model unfavorable properties (\textit{e.g.}, unfavorable physicochemical properties). Rejected samples are a strong lever for instilling expert heuristics or information about downstream properties in drug development into the protein language model. In this work, we define two types of dispreferred sequences (1) to enhance target specificity (\textit{i.e.}, binders are specific to their cognate receptors and not to other target receptors from the sample), and (2) to avoid undesirably low pI, characterized by an excessive number of negatively charged residues.

\begin{wrapfigure}{r}{0.3\textwidth}
\centering
    \includegraphics[width=0.28\textwidth]{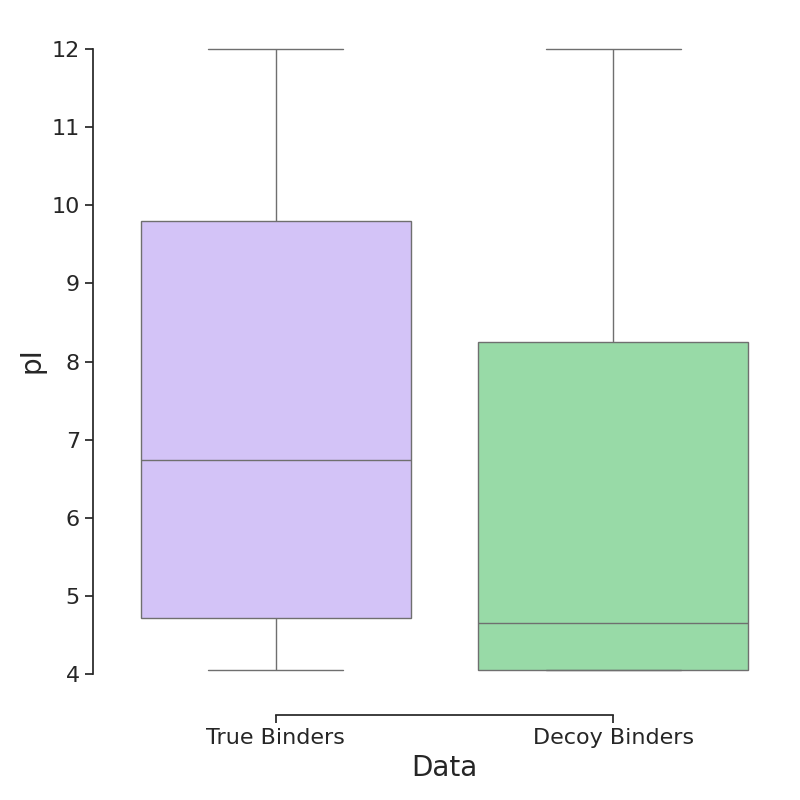}
    \caption{Statistics of isoelectric points in validation data}
    \label{fig:pI_decoys_binders_inline}
\end{wrapfigure}
\textbf{Receptor-binder dataset} We followed \citep{chen2023pepmlm} to compile protein-peptide pairs from PepNN \citep{abdin2022pepnn} and Propedia \citep{martins2023propedia} datasets. We filtered protein-peptide pairs with cutoff lengths of $500$ and $50$, respectively. We applied a homology filter with $80\%$ threshold to remove redundancies. This process led to $9,439$ pairs, involving 6,570 unique proteins and 7,557 unique peptides.

\textbf{Specificity preference dataset} We clustered the proteins and peptides separately using Mmseqs \citep{steinegger2017mmseqs2}, using a minimum sequence identity of $0.8$ and artificially constructed dispreferred protein-peptide pairs by matching ones from distant groups. A true binder peptide was assigned as a decoy dispreferred peptide to a new protein that was in a different protein-cluster. Further, to the extent possible, it was ensured that this reassigned peptide was not in the same peptide-cluster as the true binder peptides of the protein it was getting paired with. 

\textbf{Isoelectric point preference dataset} Charges on a peptide contribute to its pI (more negative charges lower its pI). We construct a dispreferred peptide from a true binding peptide by mutating $20\%$ of the residue positions to a negatively charged amino acid; the positions and the specific acid (glutamic or aspartic) are chosen randomly. Thus, for each \textit{true binder} we added a \textit{decoy binder} with lower pI, see figure \ref{fig:pI_decoys_binders_inline}.


For each protein, each true binder peptide was paired with every dispreferred decoy peptide (be it a true binder of a distant protein or a mutated form of another true binder of the same protein) to construct one training example for the DPO phase of training. This enrichment with preference datasets led to $66,898$ triplets of proteins, binders, and decoys. We held out $3,345$ triplets for testing.

\section{Results}
\label{sec:Results}

\begin{figure}[h]
\begin{center}
\subfigure[Losses during training]{
\includegraphics[width=0.48\linewidth]{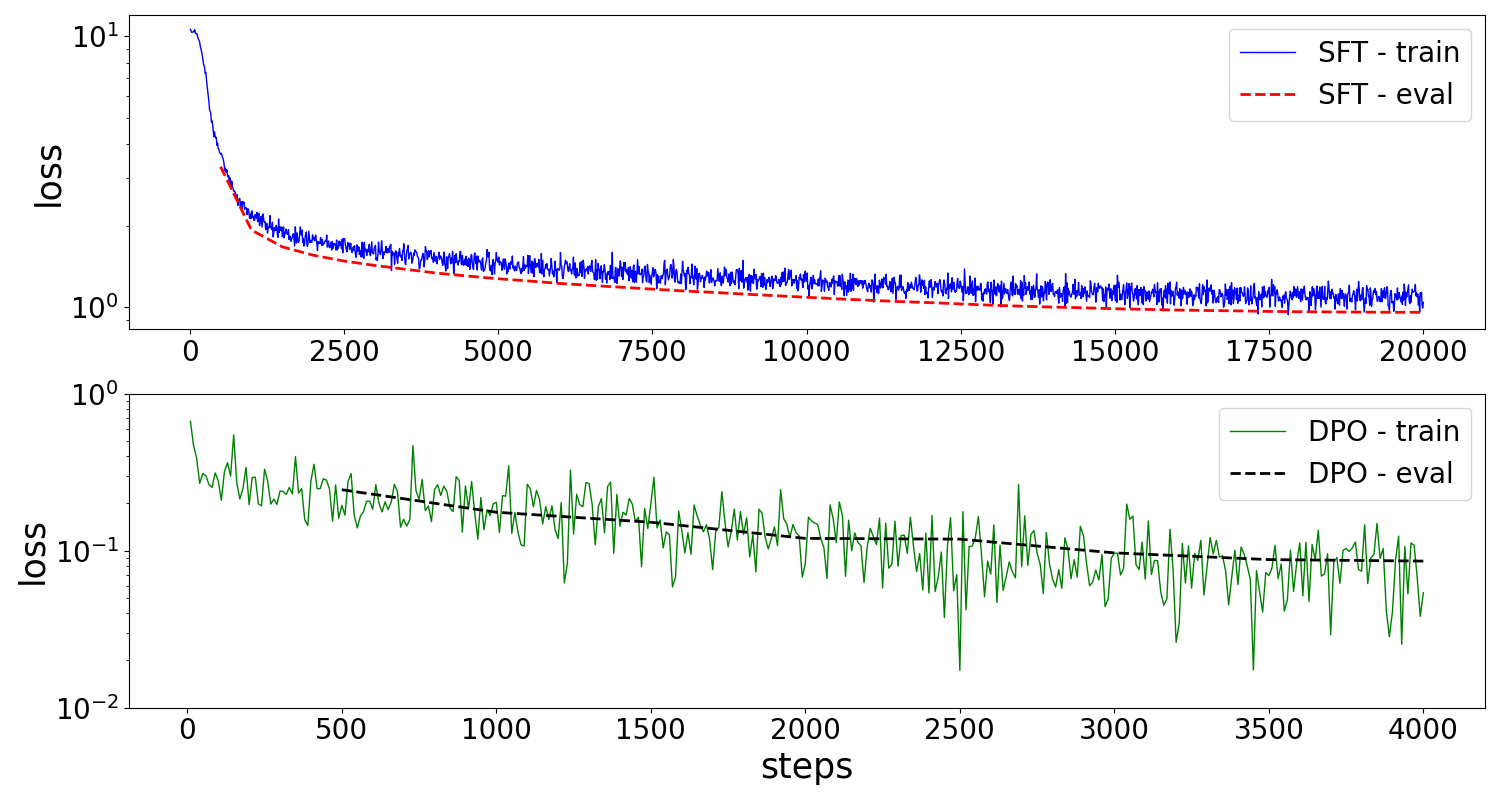}
\label{fig:losses}
}
\hfill
\subfigure[DPO rewards (chosen, rejected, margin, accuracy)]{
\includegraphics[width=0.48\linewidth]{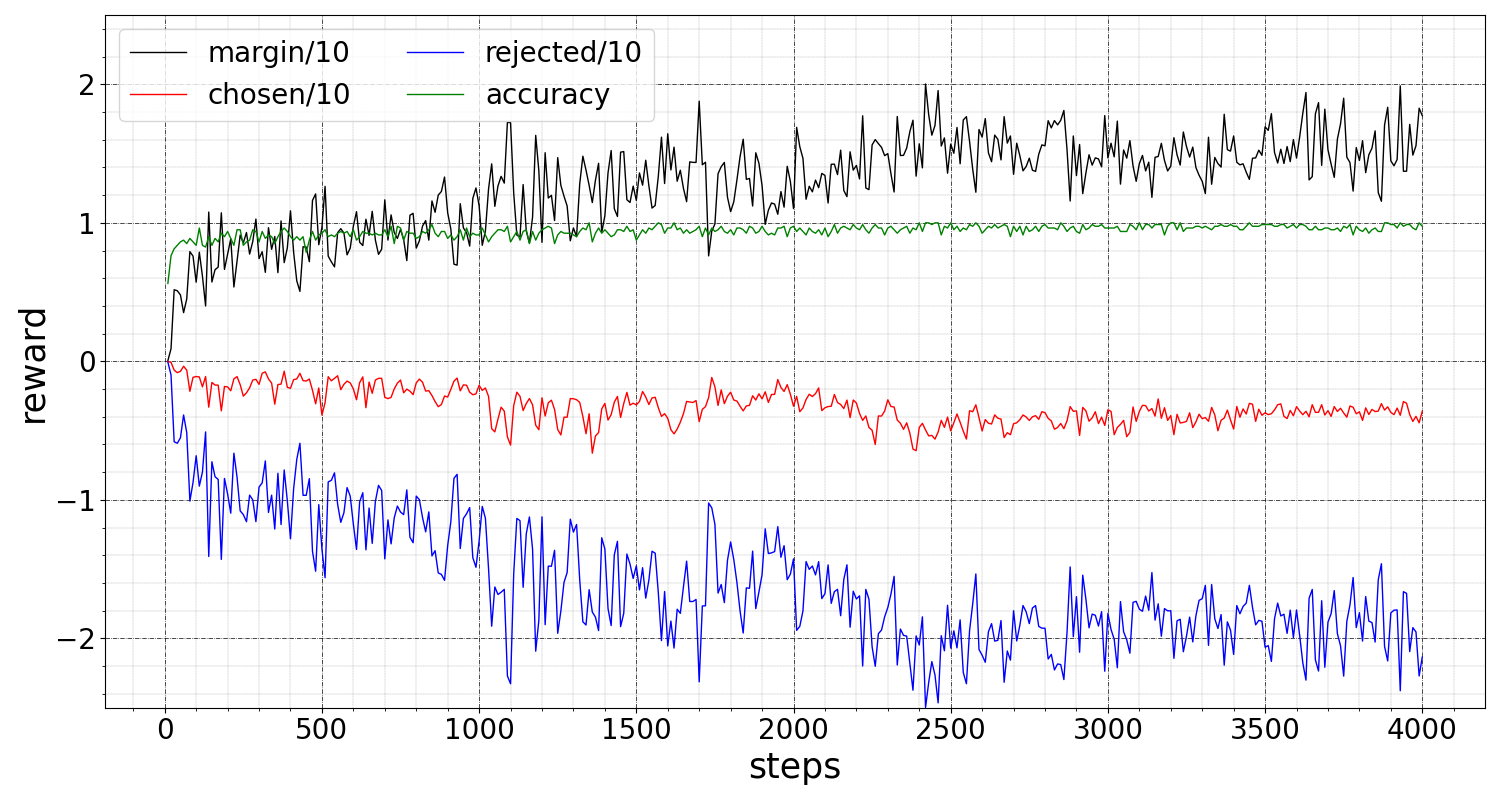}
\label{fig:rewards}
}
\end{center}\vspace{-0.2in}
\caption{Training metrics for SFT and DPO. See appendix \ref{app:dpometrics} for definition of these metrics.}
\label{fig:metrics}
\end{figure}

\textbf{Training metrics} The SFT stage instills receptor-binder completions into ProtGPT2 by training on the preferred binders, reaching a local minimum. At this state the model has maximal log probabilities for the \textit{chosen} samples, and any further optimization during DPO for discriminating the \textit{rejected} samples will inevitably degrade the log probability of the chosen data, note this is an expected outcome for any multi-objective optimization. The DPO optimization objective is maximizing the \textit{preference reward} (subject to KL divergence from the fine-tuned model), but the SFT objective is to maximize the chosen log probabilities. Figure \ref{fig:losses} illustrates that both losses decrease to the local minimum, and figure \ref{fig:rewards} shows reward metrics improve during DPO where both chosen and rejected reward log-probabilities decrease while the rejected reward decreases much faster, leading to improving margin and accuracy. Detailed definition of these metrics is given in appendix \ref{app:dpometrics}. Overall, after 24 epochs of SFT and, subsequently, 1 epoch of DPO we achieved an accuracy of $97.5\%$ and a reward margin of $17.7$. On a held-out test data accuracy and margin are $97\%$ and $15.3$, respectively. 
\textbf{Binder perplexity} Perplexity \citep{jelinek1977perplexity} is commonly used for evaluating autoregressive models. Sequence perplexity (PPL) is the exponentiated average of model negative log-likelihoods for tokens in the sequence conditioned on their previous tokens. We used beam search, top-k, and greedy sampling to generate up to 3 binders for each receptor in our test data, see appendix \ref{app:sampling}. Figure \ref{fig:ppl} compares PPLs of both SFT and DPO models and shows $50\%$ of binders exhibit a PPL less than $\sim 1.5$ and $90\%$ are below $40$. Moreover, we observe DPO does not significantly deteriorate PPLs. 
\begin{figure}[h]
\begin{center}
\subfigure[Perplexities with beam-search]{
\includegraphics[width=0.3\linewidth]{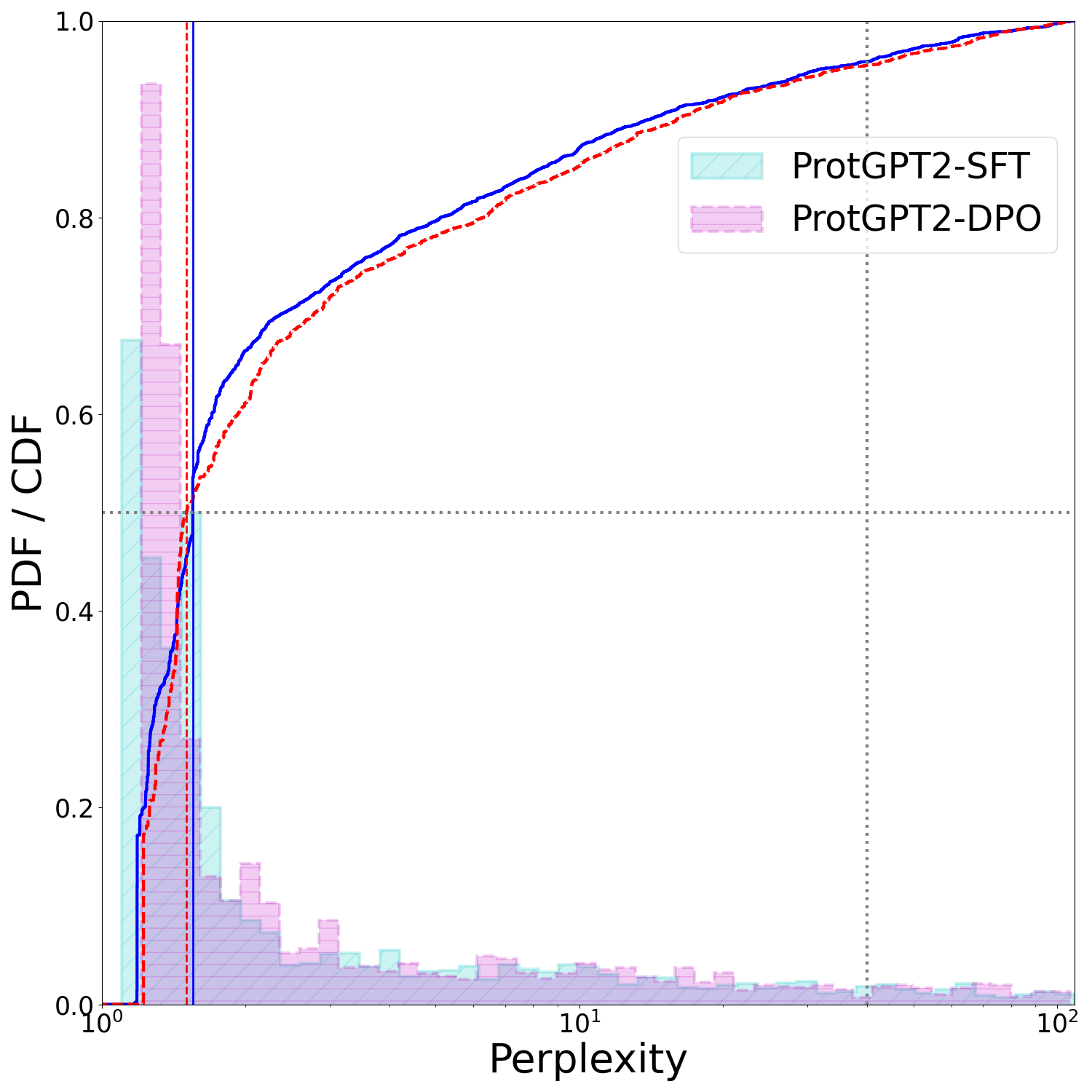}
\label{fig:ppl}
}
\hfill
\subfigure[Isoelectric points, pI]{
\includegraphics[width=0.32\linewidth]{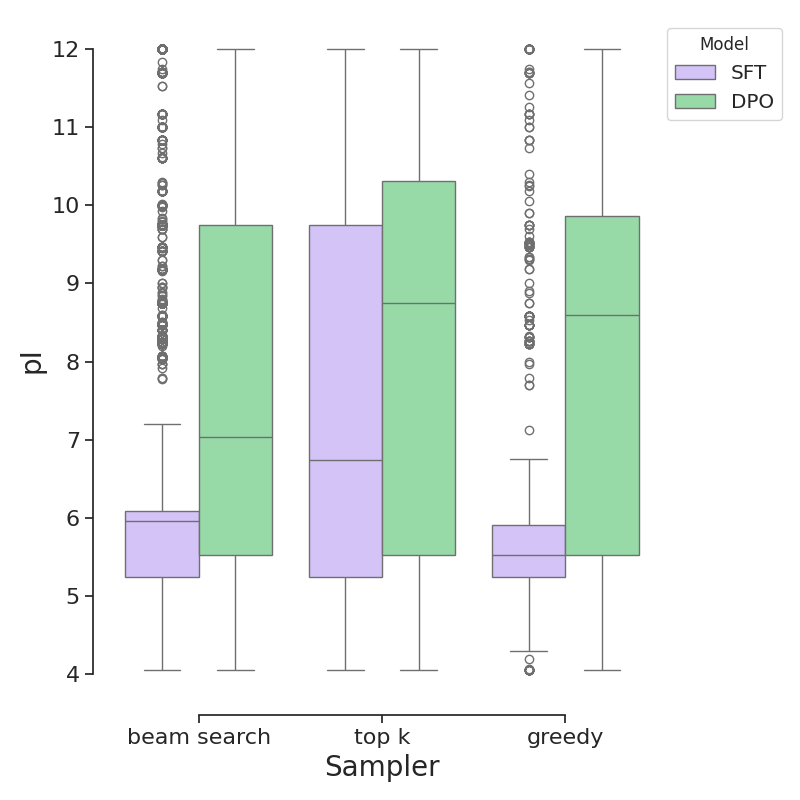}
\label{fig:pI_compare}
}\hfill
\subfigure[Alignment scores]{
\includegraphics[width=0.32\linewidth]{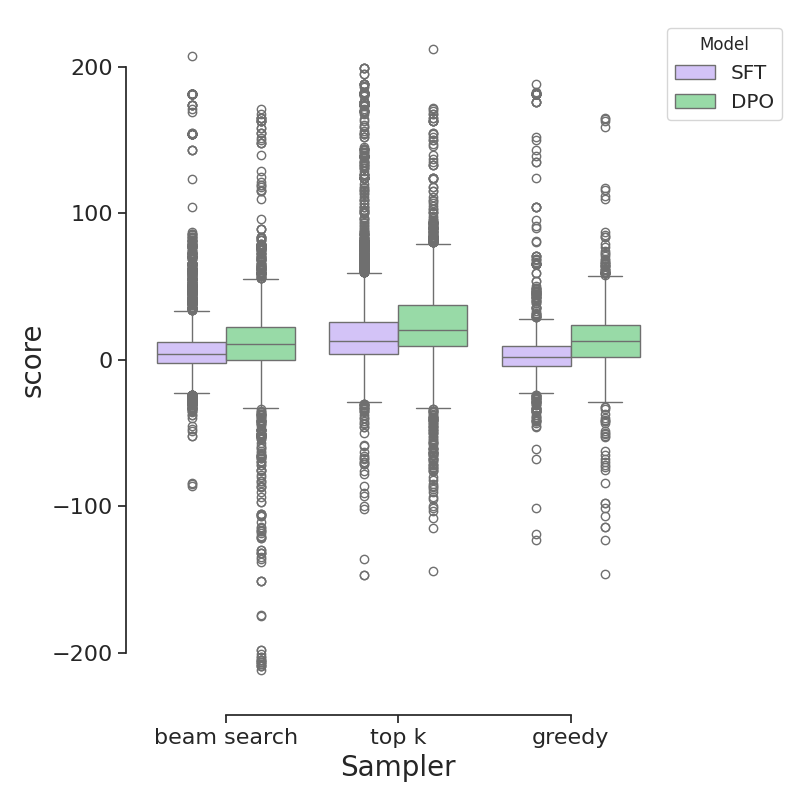}
\label{fig:score}
}
\end{center}
\vspace{-0.2in}
\caption{Generated binders by both SFT and DPO have low perplexities (left). DPO significantly improves pI (middle) and alignment scores (right)}
\label{fig:binders}
\end{figure}

\textbf{Binder pI}
Figure \ref{fig:pI_compare} shows DPO enhances the median pI values by a factor of $1.2$, $1.3$, or $1.5$ over SFT generations using beam-search, top-K, or greedy samplers, respectively. Interestingly, we observe after DPO $50\%$ of binders undergo pI improvements by a factor of $2$. 

\textbf{Binder alignment score} To examine similarity of generated binders with distributions of ground truth binders in the validation data, for each generated binder we computed the best alignment score to all positive true binders of receptors in the same cluster as its cognate protein. Figure \ref{fig:score} illustrates alignment score enhancements by DPO over the baseline SFT designs. Overall, all pI and score improvements are strong and in the expected direction in all three sampling strategies.

\section{Conclusions}
To the best of our knowledge, this is the first incorporation of DPO in protein language models for generating peptides with desirable physicochemical properties that bind to a given target protein. While the work describes adapting DPO into pLMs for peptide design, the approach is equally applicable to protein or small molecule design. Rather than having a down-stream classifier or regressor for property prediction and acceptance / rejection of a designed molecule, the framework proposed in this work affords a streamlined way to incorporate preferences ahead of time. Importantly, it opens the way to utilizing the vast amount of negative data that would have previously been deemed irrelevant during pre-training or fine-tuning. Additionally, the model can be shown dispreferred examples labeled unacceptable for reasons that may not be quantifiable by a numerical threshold on a single property (\textit{e.g.}, expert opinion, expression failure, synthetic feasibility), expanding the breadth of considerations. We anticipate that this approach will be integrated into peptide, protein and small molecule design projects in different settings, leading to an increase in efficiency of the drug discovery process by increasing the likelihood of hit molecules being translated into therapeutically viable lead molecules.

\bibliography{iclr2024_conference}
\bibliographystyle{iclr2024_conference}

\appendix

\section{Training parameters}\label{app:trainparams}
ProtGPT2 is trained with one special token \texttt{<|endoftext|>}, however to accommodate for the instruction tuning task we introduced two extra special tokens \texttt{<|im\_start|>} and \texttt{<|im\_end|>} to signify begin and end of prompts and responses. We repurposed \texttt{<|endoftext|>} as the padding token. These changes require retraining the embedding layers at the input of the network and at the model head due to enlarged embedding dimensions. Our model has $774$ million parameters.

We list the hyper-parameters used for training and model architecture in table \ref{tab:hypers}.
\begin{table}[h]
    \centering
    \caption{Hyper-parameters for training and model architecture.}
    \label{tab:hypers}
    \begin{tabular}{|l|l|}
        \hline
        \textbf{Model Architecture} & Decoder transformer with $36$ attention blocks and $774$ million parameters.\\
        & (ProtGPT2 model has $738$ million parameters, our model is larger due to \\
        & added tokens).\\
        & Embedding dimensionality is $1,280$; input/output space is $57,259$ tokens. \\
        \hline
        \textbf{Dataset} & Synthesized from $9,439$ unique protein-peptide pairs from \\
        & the Propedia and PepNN databases.\\
        & Compiled $66,898$ sequence triplets of ``protein-peptide-decoy'' for DPO.\\
        \hline
        \textbf{Hyperparameters} & Train/eval batch size per device 2/8. \\
        & $501$ warm-up steps; linear scheduler with learning rate $5\times 10^{-4}$; \\
        & AdamW optimizer.\\
        & QLoRA with $\alpha=16$, rank $r=16$, dropout is $0.05$. \\
        \hline
        \textbf{Training Duration} & Trained for $20,000$ SFT steps, $4,000$ DPO steps. \\
        \textbf{Hardware} & 2 NVIDIA RTX A6000 GPUs. \\
        \textbf{Training Time} & $\sim 7.5$ hours ($24.36$ epochs) for instruction fine-tuning, and\\
        & $\sim 1.3$ hour for DPO (1 epoch). \\
        \hline
        \textbf{Initialization} & Initialized with pre-trained weights from ProtGPT2 trained on 50 million\\
        & sequences from UniProt database. \\
        \hline
    \end{tabular}
\end{table}

\section{DPO metrics}\label{app:dpometrics}
Below are definitions for the reported reward metrics for DPO:
\begin{itemize}
    \item $\displaystyle chosen\ reward=\E_{(\rx,\ry_w,\ry_l)\sim\mathcal{D}} \big[\beta \log \frac{\pi_\theta(y_w|x)}{\pi_{ref}(y_w|x)}\big]$, quantifying the mean difference between the log probabilities of the policy model and the reference model on the chosen responses. 
    \item $\displaystyle rejected\ reward=\E_{(\rx,\ry_w,\ry_l)\sim\mathcal{D}} \big[\beta\log \frac{\pi_\theta(y_l|x)}{\pi_{ref}(y_l|x)} \big]$: quantifying the mean difference between the log probabilities of the policy model and the reference model on the rejected responses.
    \item $\displaystyle accuracy$: mean frequency of chosen rewards being greater than the rejected rewards.
    \item $\displaystyle margin$: the mean difference between the chosen and rejected rewards.
\end{itemize}

\section{Sampling strategies}
\label{app:sampling}
\begin{figure}[h]
\begin{center}
\subfigure[Top-K sampling with top 3 sequences per each prompt, with $\rm top\_k=950$ and $\rm top\_p=0.85$ and $\rm temperature=0.7$. Median perplexities are $\rm SFT=1.67$ and $\rm DPO=1.77$.]{
\includegraphics[width=0.3\linewidth]{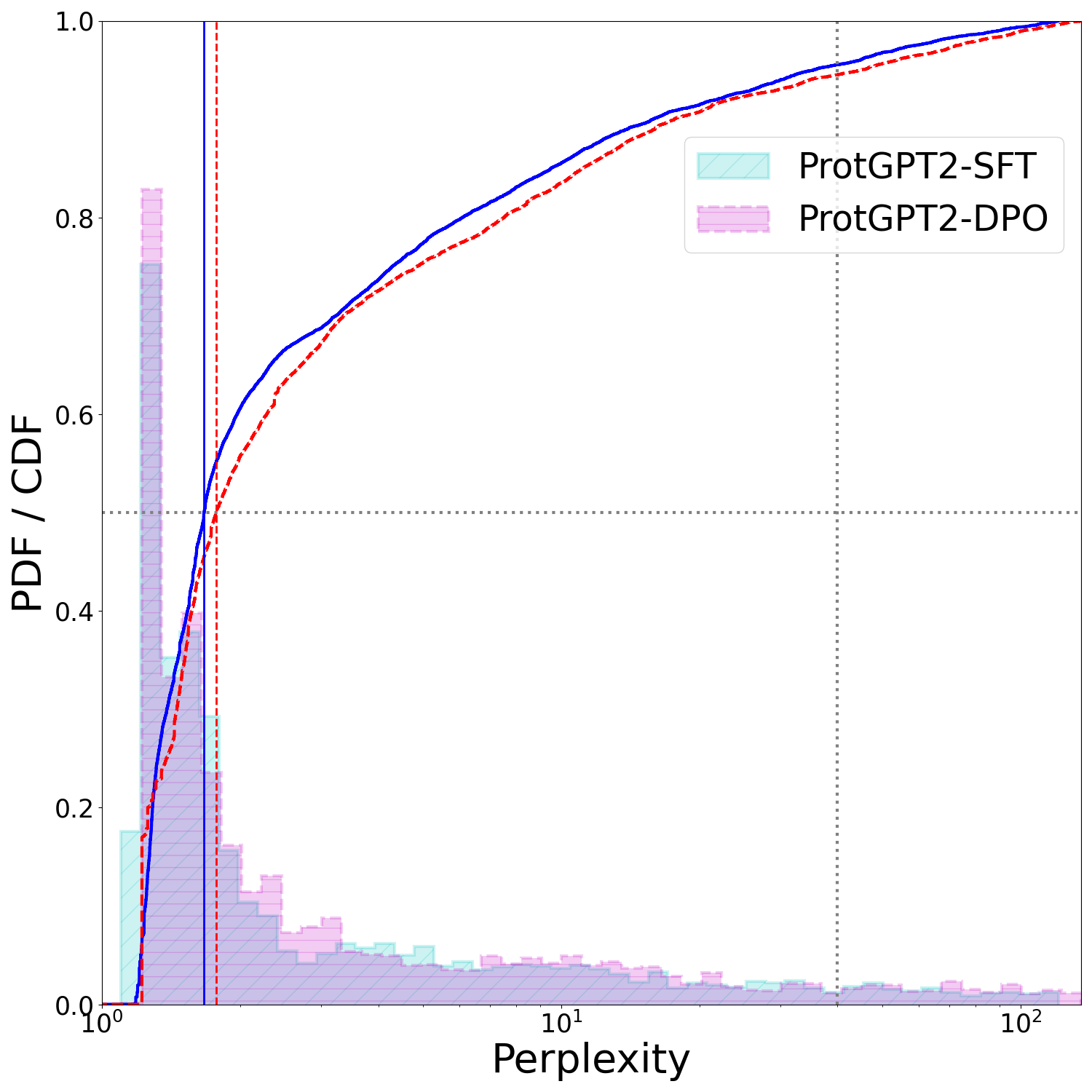}
\label{fig:topk}
}\hfill
\subfigure[Beam-search sampling with top 3 sequences out of 20 beams per each prompt. Median perplexities are $\rm SFT=1.67$ and $\rm DPO=1.50$.]{
\includegraphics[width=0.3\linewidth]{figs/perplexity_pairs_cdf_beam_search_beam_search_sampling.png}
\label{fig:beamsearch}
}\hfill
\subfigure[Greedy sampling. Median perplexities are $\rm SFT=1.67$ and $\rm DPO=1.90$.]{
\includegraphics[width=0.3\linewidth]{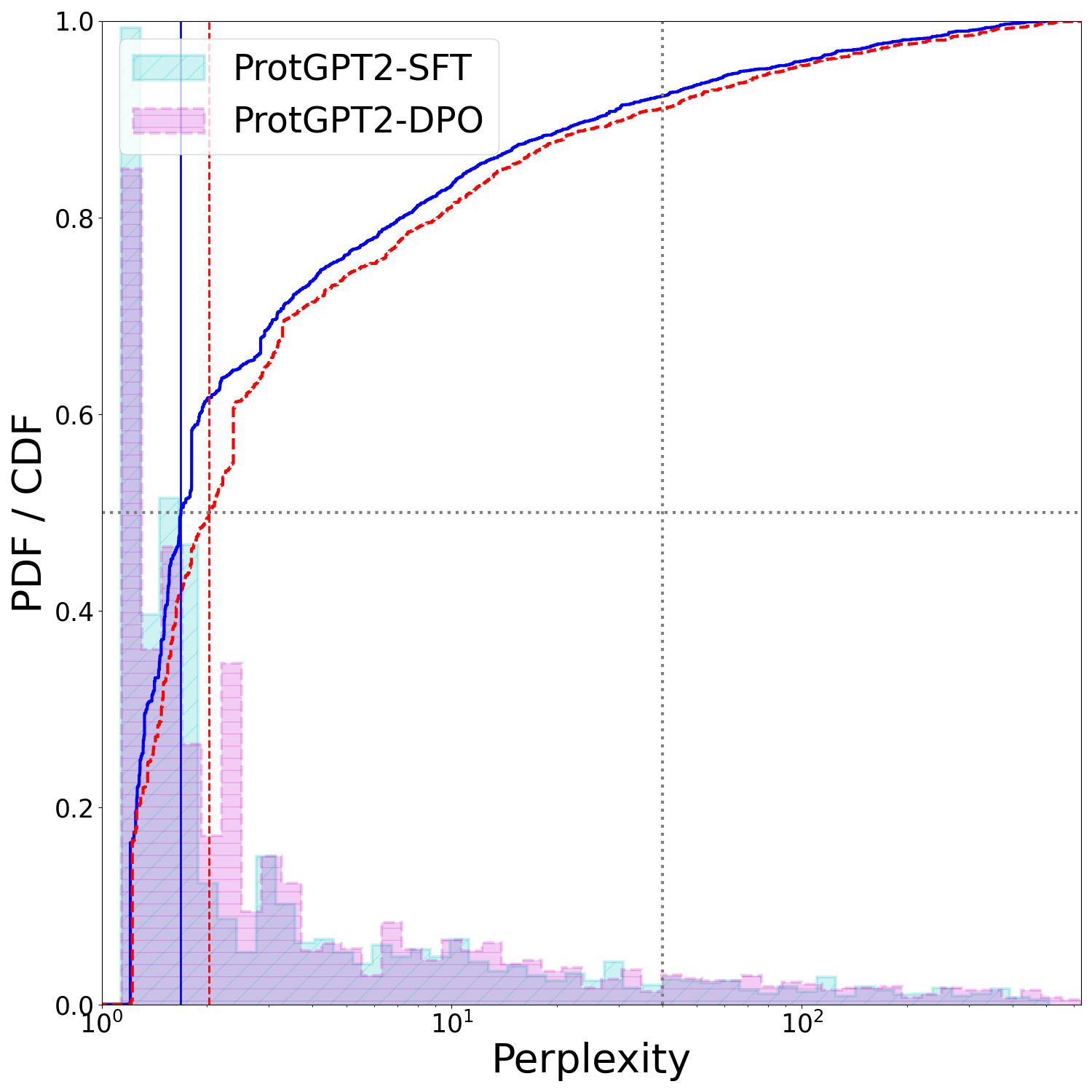}
\label{fig:greedy}
}
\end{center}
\caption{Probability (and cumulative) distribution functions for perplexities computed with different sampling strategies. Receptors from a held-out validation set were used to prompt the models for binder designs.}
\label{fig:samplingperplexity}
\end{figure}

 Different sampling strategies generate sequences with varying levels of model uncertainty. Here we experimented influences of different sampling strategies on induced sequence perplexities. A higher perplexity implies higher model uncertainty. We experimented with different sampling strategies including top-k \citep{fan2018hierarchical} (with top-p), greedy, and beam search. Beam-search generated lowest binder perplexities when conditioned on receptor sequences; see figures \ref{fig:samplingperplexity}.


\end{document}